# Large Diamagnetism of $AV_2Al_{20}$ (A = Y and La)

Atsushi ONOSAKA, Yoshihiko OKAMOTO, Jun-ichi YAMAURA,
Takahiro HIROSE, and Zenji HIROI[*]

*Institute for Solid State Physics, University of Tokyo, 5-1-5 Kashiwanoha, Kashiwa, Chiba 277-8581, Japan*

We find anomalously large diamagnetic responses in the cage compounds $AV_2Al_{20}$ where A = Y and La, not A = $Al_{0.3}$, $Sc_{0.4}$, and Lu, despite the apparent similarities in crystal and electronic structures among these compounds. The magnetic susceptibilities of the Y and La compounds become -1.94 and -7.44 × $10^{-4}$ $cm^3$ $mol^{-1}$ at 10 K, respectively, the latter of which corresponds to approximately one-quarter of that of bismuth, a well-known diamagnetic material, in terms of unit volume. The origin is not clear but may be related to a specific evolution in the band structure, as the diamagnetic response increases with increasing lattice constant.

[*]E-mail address: hiroi@issp.u-tokyo.ac.jp

KEYWORDS: diamagnetism, cage compound, $YV_2Al_{20}$, $LaV_2Al_{20}$

The magnetism of compounds is generally governed by two degrees of freedom of an electron, i.e., spins and orbitals. Most metallic compounds exhibit paramagnetic responses in total, because a paramagnetic response from spins, which is called the Pauli paramagnetism, exceeds a diamagnetic response from orbitals, called Landau diamagnetism or orbital diamagnetism; the former is three times larger in magnitude than the latter in the free-electron model. However, this is not always the case for some semimetallic compounds having conduction electrons with very small effective masses, such as graphite[1] and bismuth;[2] the magnitude of Landau diamagnetism is inversely proportional to the effective mass. Graphite and bismuth show large diamagnetic responses: $\chi$ = -2.58 × $10^{-4}$ and -2.80 × $10^{-4}$ $cm^3$ $mol^{-1}$ at room temperature, respectively. The $\chi$ of graphite is highly anisotropic, is large in a magnetic field perpendicular to the basal plane containing a large electron orbital, and shows a characteristic temperature dependence, increasing in magnitude and saturating at -3.29 × $10^{-4}$ $cm^3$ $mol^{-1}$ at low temperature.[3] These features are understandable in terms of the simple band structure of graphite. In contrast, the $\chi$ of bismuth is much larger than expected from a free-electron model based on five Fermi surfaces and is interpreted as coming from an interband effect in the presence of a strong spin-orbit coupling.[4,5] Recently, orbital diamagnetism has been reviewed in Dirac fermion systems like graphene[6] and a molecular conductor,[7] because they have massless electrons that can give rise to a huge enhancement in orbital diamagnetism. There is another diamagnetic contribution in compounds, i.e., core diamagnetism or Lamour diamagnetism. It is a response against a magnetic field from core electrons in a closed shell of an ion and is always much weaker than the above-mentioned magnetic responses; the magnitude tends to increase with increasing atomic number Z. Moreover, it is temperature-independent.

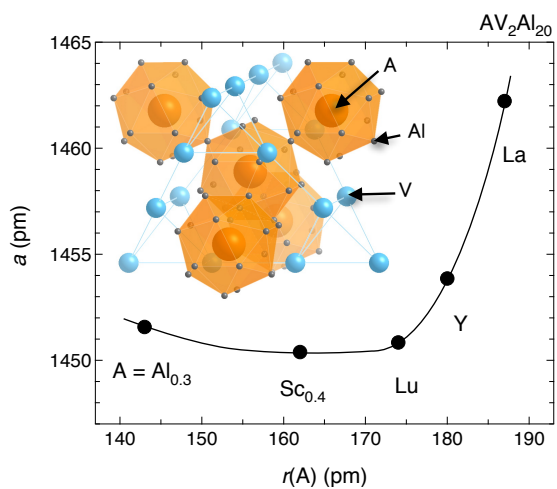

**Fig. 1.** (Color online) Cubic lattice constant $a$ of $AV_2Al_{20}$ at room temperature as a function of the radius of A atoms in metal $r$(A). The line on the data is a visual guide. The inset depicts the $CeCr_2Al_{20}$-type crystal structure with the space group $Fd$-$3m$ (origin choice 2). The A atom at the 8$a$ position is confined in a cage made of four 16$c$ Al and twelve 96$g$ Al atoms. The sets of A and V (16$d$) atoms form diamond and pyrochlore lattices, respectively.

A family of ternary intermetallic compounds with the general formula $AB_2C_{20}$ crystallize in the cubic $CeCr_2Al_{20}$-type structure of space group $Fd$-$3m$.[8] Transition metal B atoms occupy the 16$d$ position and form a corner-sharing tetrahedral network called the pyrochlore lattice, and A atoms at the 8$a$ position that are located inside high-symmetry cages made of C atoms form a diamond lattice, as shown in Fig. 1. Many intriguing properties have been found in the family, such as a large electronic heat capacity in $YbCo_2Zn_{20}$,[9] and superconductivity and quadrupole order in $PrIr_2Zn_{20}$[10] and $PrTi_2Al_{20}$[11]. Very recently, $AV_2Al_{20}$ with A = $Al_x$ or $Ga_x$ has been studied to reveal the rattling of A atoms and its effect on superconductivity;[12,13] Al and Ga atoms are relatively small so that they can rattle inside the cage.[14] These compounds are stable only when the A site is partially occupied, e.g. $x \sim 0.3$ for A = $Al_x$.

In the course of our study of rattling in $AV_2Al_{20}$, we prepared isomorphic compounds with A = $Sc_{0.4}$, Y, La, and Lu. The $Sc_{0.4}$ compound exhibits less rattling and the others show no rattling because the A atoms are large enough to fit the cage. In fact, as shown in Fig. 1, the cubic lattice constant at room temperature remains intact or even decreases slightly with increasing the atomic radius of A atoms in metal from A = Al to Sc, while it increases for A = Y, further increasing for La; the intrinsic size of the cage may be approximately 175 pm for A = Lu. Surprisingly, we have observed large diamagnetic responses only for the Y and La compounds in spite of the apparent similarities in the crystal and electronic structures among the series. Particularly, $LaV_2Al_{20}$ exhibits a diamagnetic response of $-7.44 \times 10^{-4}$ cm$^3$ mol$^{-1}$ at 10 K, which corresponds to approximately one-quarter of that of bismuth in terms of unit volume. The origin of the large diamagnetism is not clear at the moment but may be related to a subtle evolution in the band structure as the lattice constant increases.

Five polycrystalline samples of $AV_2Al_{20}$ with A = $Al_{0.3}$, $Sc_{0.4}$, Y, La, and Lu were prepared from A elements, aluminum, and vanadium in appropriate molar ratios, as reported previously.[12] First, a composite was melted at a high temperature in an arc-melt furnace to obtain a uniform mixture. After sealing in an evacuated quartz ampoule, the mixture was annealed at 650-800 ºC. Powder X-ray diffraction measurements revealed that monophasic samples were obtained. The lattice constant of the cubic unit cell is shown in Fig. 1 and Table I. The chemical compositions were examined by energy-dispersive X-ray analysis in scanning electron microscopy, which confirmed that the intended compositions were nearly retained in the products. Magnetic susceptibility and heat capacity were measured in a magnetic property measurement system and a physical property measurement system (Quantum Design), respectively.

We have measured magnetization as a function of magnetic field at 10, 100, and 300 K for all the polycrystalline samples of $AV_2Al_{20}$. Figure 2(a) shows typical $M$-$H$ curves measured at 100 K upon increasing and decreasing magnetic field between 0 and 70 kOe. The $Al_{0.3}$, $Sc_{0.4}$, and Lu compounds exhibit paramagnetic responses with positive slopes of similar magnitudes, while the Y and La compounds are diamagnetic with negative slopes, particularly large for the La compound. All the $M$-$H$ curves are linear at high magnetic fields. However, there is a weak additional component at low fields, particularly discernible for the La compound in Fig. 2(a), which already saturates at low magnetic fields at approximately 3 kOe. Thus, the $M$-$H$ curves at large fields can be fitted to $M = M_0 + \chi H$, in which $M_0$ and $\chi$ are an additional magnetization at $H$ = 0 and the magnetic susceptibility at the temperature used, respectively. The estimated $M_0$'s are 0.97 and 3.00 G cm$^3$ mol$^{-1}$ for the Lu and La compounds, respectively, and are negligible for the other compounds. The origin of $M_0$ is not clear but may be the small amount of ferromagnetic impurities such as a compound containing iron, because $M_0$ is almost constant below room temperature (its Curie temperature must be higher than room temperature) and also because it changes from sample to sample even with $\chi$ unchanged. Thus, intrinsic $\chi$'s at 10 and 100 K have been estimated by subtracting $M_0$ and are listed in Table I.

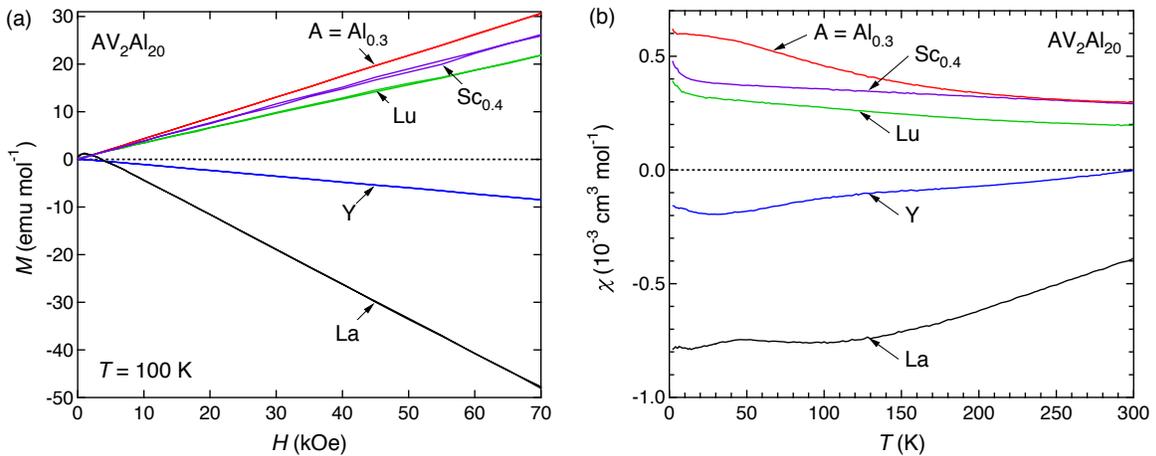

**Fig. 2.** (Color online) (a) Magnetizations $M$ of five polycrystalline samples of $AV_2Al_{20}$ (A = $Al_{0.3}$, $Sc_{0.4}$, Y, La, and Lu) measured at 100 K upon increasing and decreasing magnetic field. (b) Temperature dependences of magnetic susceptibility $\chi$ measured at $H$ = 10 or 50 kOe after correction of ferromagnetic contributions observed in $M$-$H$ curves in (a).

Figure 2(b) shows the temperature dependences of $\chi$ obtained by subtracting the ferromagnetic components estimated at 100 K. First, let us focus on the $\chi$ of LuV$_2$Al$_{20}$. It is positive and slightly increases upon cooling, reaching 3.14 × 10$^{-4}$ cm$^3$ mol$^{-1}$ at 10 K; the small upturn below 10 K may be due to nearly free spins around defects or in impurities. The weak temperature dependence of $\chi$ may reflect the fact that the Fermi level is located just around a sharp peak in the density of states (DOS): the effective DOS becomes larger upon cooling with decreasing thermal smearing; thus, $\chi$ gradually decreases at low temperatures. As will be mentioned later, this $\chi$ is dominated by the Pauli paramagnetism. On the other hand, the Sc$_{0.4}$ compound shows a slightly larger $\chi$ with a similar temperature dependence, and the $\chi$ of the Al$_{0.3}$ compound merges with that of Sc$_{0.4}$ above 200 K. Thus, the $\chi$'s of these three compounds are similar to each other and may be reasonably understood.

In contrast, the $\chi$'s of the Y and La compounds are anomalous. The $\chi$ of the Y compound is nearly zero at 300 K, becomes negative upon cooling, and reaches -1.94 × 10$^{-4}$ cm$^3$ mol$^{-1}$ at 10 K. The La compound shows a similar temperature dependence with a much larger diamagnetic response, reaching -7.44 × 10$^{-4}$ cm$^3$ mol$^{-1}$ at 10 K. Moreover, a broad hump is discernible at ~ 45 K in the La compound. Apparently, these variations in magnetic susceptibility should be ascribed to the difference in the A elements, although the number of A atoms is merely 1/23 in the formula unit of AV$_2$Al$_{20}$. Note that, at least, the Lu, Y, and La compounds have the same electron filling, so that the contributions of conduction electrons may not be so different.

Heat capacity measurements have been performed to extract information on the electronic structures of the compounds. Figure 3 shows the $C/T$ vs $T^2$ plots below 10 K. A linear relation is observed below ~7 K for the Lu, Y, and La compounds, as expected for normal metallic compounds. A fit to the equation $C = \gamma T + \beta T^3$, where $\gamma$ is the Sommerfeld coefficient and $\beta$ gives the Debye temperature $\Theta_D$ as $\Theta_D = [1943.7 \times 23 / \beta]^{(1/3)}$, yields $\gamma$ and $\Theta_D$ as listed in Table I. In the Al$_{0.3}$ and Sc$_{0.4}$ compounds, there are large contributions of the low-energy rattling vibrations of Al atoms at low temperatures. Taking this into account, $\gamma$ was decided in a previous study.[12] As a result, the $\gamma$'s of all these compounds are moderately large, 20-33 mJ K$^{-2}$ mol$^{-1}$, and decrease gradually with increasing $r$(A), as shown in Fig. 4(a). Note that the $\gamma$'s of the Lu and Y compounds are equal, suggesting basically the same DOS at the Fermi level, in spite of the fact that they show magnetic susceptibilities with opposite signs.

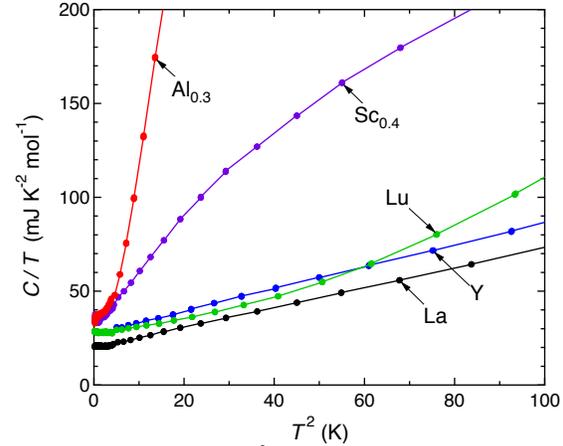

**Fig. 3.** (Color online) $T^2$ plot of heat capacity divided by temperature $C/T$ for polycrystalline samples of AV$_2$Al$_{20}$ (A = Al$_{0.3}$, Sc$_{0.4}$, Y, La, and Lu) measured below 10 K upon cooling in zero magnetic field.

Let us discuss the magnetic susceptibility of AV$_2$Al$_{20}$. In general, the $\chi$ of a metallic compound is given by the summation of three components: $\chi = \chi_P + \chi_L + \chi_{core}$, where $\chi_P$, $\chi_L$, and $\chi_{core}$ represent the Pauli paramagnetism, Landau or orbital diamagnetism, and the magnetisms from core electrons, respectively. The first and second terms are from conduction electrons: $\chi_e = \chi_P + \chi_L$. The last term includes two contributions: Lamour diamagnetism and Van Vleck paramagnetism. The Lamour diamagnetism of AV$_2$Al$_{20}$ is given by the summation of all atomic contributions and evaluated on the basis of the literature values of -2.5, -4, -7.7, -24, -38, and -17 × 10$^{-6}$ cm$^3$ mol$^{-1}$ for Al$^{3+}$, Sc$^{3+}$, V$^{5+}$, Y$^{3+}$, La$^{3+}$, and Lu$^{3+}$, respectively.[15] For example, $\chi_{core}$ = -0.66 and -1.03 × 10$^{-4}$ cm$^3$ mol$^{-1}$ for Al$_{0.3}$V$_2$Al$_{20}$ and LaV$_2$Al$_{20}$, respectively (see Table I for the other compounds). On the other hand, the Van Vleck paramagnetism is difficult to estimate and is often relatively small except for those of the iron group elements. Here, we ignore this contribution, because that of Al, the main element of AV$_2$Al$_{20}$, must be negligible. Thus, $\chi_P + \chi_L$ can be estimated after the subtraction of $\chi_{core}$. As mentioned in the introduction, the positive $\chi_P$ is larger than the negative $\chi_L$ in magnitude in most metals, so that $\chi_e$ becomes positive.

The observed negative $\chi$'s for the Y and La compounds suggest that $\chi_L$ is unusually enhanced for some reason.

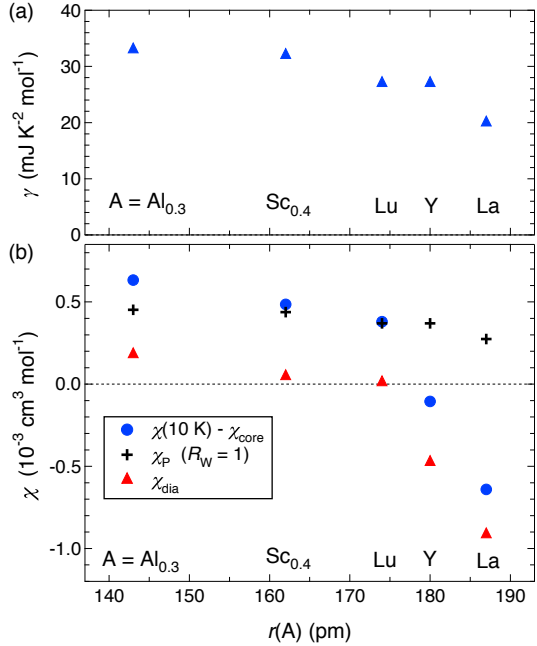

**Fig. 4.** (Color online) Sommerfeld coefficient $\gamma$ (a) and magnetic susceptibility $\chi$ (b) as functions of the metallic radius of the A elements $r$(A). Plotted in (b) are the experimental $\chi$ at 10 K after subtraction of $\chi_{core}$, $\chi_P$ estimated from $\gamma$ assuming $R_W = 1$, and $\chi_{dia} = \chi(10\,K) - \chi_{core} - \chi_P$.

Among all the compounds in the series, LuV$_2$Al$_{20}$ looks quite normal. $\chi - \chi_{core}$ at 10 K is $3.80 \times 10^{-4}$ cm$^3$ mol$^{-1}$ and the Wilson ratio $R_W$, which is given by $(\pi^{2/3})(k_B/\mu_B)^2(\chi_P/\gamma)$, is close to unity: $R_W = 1.03$. This means that $\chi_e$ is dominated by $\chi_P$ with negligible $\chi_L$ and also that electron correlation is weak in this compound. Thus, LuV$_2$Al$_{20}$ is a conventional metal and may provide us with a standard for examining the properties of the other compounds. The $\chi$ of Sc$_{0.4}$V$_2$Al$_{20}$ is slightly larger than that of LuV$_2$Al$_{20}$, reflecting a larger $\gamma$, $R_w = 1.11$. The $\chi$ of the Al$_{0.3}$ compound coincides with that of Sc$_{0.4}$ above 200 K, which is consistent with the similar $\gamma$ values of these compounds. Therefore, both the magnetic susceptibility and heat capacity of all three compounds are consistent with each other and quite conventional.

In sharp contrast, the magnetic susceptibilities of Y and LaV$_2$Al$_{20}$ are distinguished. Their $\chi_P$'s are estimated to be 3.70 and $2.74 \times 10^{-4}$ cm$^3$ mol$^{-1}$, respectively, assuming $R_w = 1$. Obviously, there are large additional diamagnetic contributions: $\chi_{dia} = \chi(10\,K) - \chi_P - \chi_{core}$. $\chi_{dia}$ is shown as a function of the metallic radius of A elements in Fig. 4(b). It increases in magnitude for Y and further for La, reaching a large value of $-10^{-3}$ cm$^3$ mol$^{-1}$, which corresponds to approximately one-quarter of that of bismuth in terms of unit volume.

It is reasonable to ascribe the origin of the large $\chi_{dia}$ to orbital magnetism: there must be a subtle but significant change in the band structure. Since $\gamma$ decreases slightly with increasing $r$(A) but still remains large, heavy bands commonly exist. Plausibly, an additional very light band appears for the Y and La compounds, which gives rise to a large orbital diamagnetism. However, it would be unrealistic to assume the existence of such extremely light carriers as in graphite for the present compounds containing heavy metals and 3$d$ electrons. We think that, in addition to the heavy bands, light electron and hole bands approach the Fermi level, as the lattice constant increases, which may be effective for enhancing diamagnetism through an interband effect as observed in bismuth. Band structure calculations and further experiments are in progress to reveal the origin of the large diamagnetism particularly in LaV$_2$Al$_{20}$.

In summary, we have observed anomalously large diamagnetic responses from the cage compounds YV$_2$Al$_{20}$ and LaV$_2$Al$_{20}$, but not from isomorphic compounds containing Al$_{0.3}$, Sc$_{0.4}$, and Lu at the A site of AV$_2$Al$_{20}$. It is suggested that light bands suddenly appear for the Y and La compounds as the lattice constant increases, in addition to common heavy bands. The relevance of this diamagnetic response to the diamagnetism of bismuth will be intriguing to investigate in future studies.

**Acknowledgments** We are grateful to H. Harima and Y. Fuseya for helpful discussions. This work was supported by a Grant-in-Aid for Scientific Research on Priority Areas "Heavy Electrons" (No. 23102704) provided by MEXT, Japan.

**Table I.** Various lattice and electronic parameters for $AV_2Al_{20}$ (A = $Al_{0.3}$, $Sc_{0.4}$, Lu, Y, and La). Magnetic susceptibilities are given in $10^{-4}$ cm$^3$ mol$^{-1}$.

| $AV_2Al_{20}$ | A = $Al_{0.3}$ | $Sc_{0.4}$ | Lu | Y | La |
|---|---|---|---|---|---|
| $r(A)$ [pm] | 143 | 162 | 174 | 180 | 187 |
| $a$ [pm] | 1451.57 | 1450.39 | 1450.84 | 1453.86 | 1462.22 |
| $\Theta_D$ [K] | - | - | 470 | 420 | 430 |
| $\gamma$ [mJ K$^{-2}$ mol$^{-1}$] | 33 | 32 | 27 | 27 | 20 |
| $\chi$(10 K) | 5.67 | 4.18 | 3.14 | -1.94 | -7.44 |
| $\chi$(100 K) | 4.38 | 3.69 | 2.97 | -1.24 | -7.30 |
| $\chi_{core}$ | -0.66 | -0.67 | -0.66 | -0.89 | -1.03 |
| $\chi$(10 K) - $\chi_{core}$ | 6.33 | 4.85 | 3.80 | -1.05 | -6.41 |
| $\chi_P$ ($R_W = 1$) | 4.52 | 4.38 | 3.70 | 3.70 | 2.74 |
| $\chi_{dia}$ | 1.81 | 0.47 | 0.10 | -4.75 | -9.15 |